\documentclass{PoS}

\title{Space-VLBI with RadioAstron: new correlator capabilities at MPIfR}

\ShortTitle{Space-VLBI with RadioAstron}

\author{\speaker{G. Bruni$^{1}$}, J. M. Anderson$^{1,2}$, W. Alef$^{1}$, A. Lobanov$^{1,3}$, J. A. Zensus$^{1}$  \\%\thanks{}\\
        $^{1}$Max-Planck-Institut f\"ur Radioastronomie, Auf dem H\"ugel 69, 53121 Bonn, Germany \\
	  $^{2}$Deutsches GeoForschungsZentrum GFZ, Telegrafenberg A6, 14473 Potsdam, Germany\\
        $^{3}$Institut f\"ur Experimentalphysik, Universit\"at Hamburg, Luruper Chaussee 149, 22761 Hamburg, Germany\\
        E-mail: \email{bruni@mpifr-bonn.mpg.de}}

\abstract{DiFX is a correlator for VLBI data based on the FX architecture (first Fourier transform and then cross-multiply). DiFX is a free licensed software written in C++, developed and maintened by an international group of programmers. \\
A new DiFX version ({\it{dra}}) has been developed at Max-Planck-Institut f\"ur Radioastronomie (MPIfR), in order to manage the correlation of a space-based antenna with ground stations.\\
The {\it{dra}} version is running on the High Performance Computer cluster (HPC) in Bonn, and it is used for the data processing of the three AGN imaging RadioAstron Key Science Projects ongoing, based at the MPIfR.
}

\FullConference{12th European VLBI Network Symposium and Users Meeting,\\
		7-10 October 2014\\
		Cagliari, Italy}

\begin{document}

\section{The RadioAstron mission}

RadioAstron is an international space VLBI (SVLBI) mission led by the Astro Space Center (ASC, Moscow) and utilizing a 10-meter space radio telescope (SRT) on board of the Russian satellite {\it{Spektr-R}} which was launched in July 2011 [2]. Since beginning of 2012, it is performing interferometric observations with ground-based telescopes from all over the world. Key science projects as well as PI-driven projects have been observed until now during Announcement of Opportunity 1 (AO-1, July 2013 - July 2014), and AO-2 (July 2014 - July 2015). The MPIfR plays a leading role in three KSPs aimed to study structure, evolution, and polarisation properties of extragalactic jets, taking advantage of the extreme angular resolution offered by RadioAstron. Since summer of 2013, the Bonn correlator is processing data from these projects: during AO-1, 13 experiments were observed for the three KSPs.

The correlation of the collected data is carried out at the High Performance Computer (HPC) cluster, based at MPIfR. It features 480 cores, 13 RAIDs (650 Tb of storage capability), 20 Gbps infiniband interconnection, and 15 Mark5 units for diskpacks playback. The HPC cluster is currently used also for correlation of mm-VLBI, and geodetic projects.

\section{The \emph{dra} version of DiFX}

%The DiFX software (Deller et al. 2007) correlator has been upgraded by the MPIfR in Bonn to enable space-VLBI (SVLBI) correlation, addressing specific aspects of data and telemetry formats of RadioAstron.

As a contribution to the RadioAstron mission, the MPIfR devoted
efforts to enhance the DiFX correlator [1] to enable space-VLBI (SVLBI) correlation
including the RadioAstron telescope.
Although the RadioAstron mission continues to develop its own software
correlator in the Astro Space Center (ASC) Moscow, the mission desired
having an independent correlator available to be able to check the ASC
correlator results.  Furthermore, enhancing the DiFX correlator for
SVLBI enables the MPIfR to perform correlation of RadioAstron
experiments for MPIfR science projects as a contribution to the
RadioAstron mission, and since DiFX is widely used by the VLBI
community, RadioAstron correlation could in principle be performed by
any other DiFX correlator group around the world.  

During initial data format testing in 2012, D. Graham (MPIfR) wrote software to
convert the RadioAstron raw-voltage data format (RDF) to the Mark5B
file format based on reverse engineering the sample data obtained from
the RadioAstron mission.  This enables the RadioAstron data to be
converted into a format supported by DiFX. J. M. Anderson later modified this conversion software to correct
for changes made to the format since 2010 and to incorporate
documentation on the data format obtained in 2012.

Initial efforts to modify the DiFX correlator software itself were
started in 2011. The most significant part of this effort
involved modifying the delay model server (Calc, from the Calc/Solve
package\footnote{http://gemini.gsfc.nasa.gov/solve/}) to be able to
calculate delay information for telescopes with arbitrary coordinates
and velocities (that is, not fixed on the ground), and changes to the
DiFX metadata system to deal with the changing position and velocity
of the spacecraft as a function of time.  The delay model was also
modified to correct for (general) relativistic effects as the highly
elliptical orbit of the spacecraft results in large changes in
velocity and gravitational potential compared to the terrestrial
frame.  Because of hardware issues in the equipment used for receiving
and recording the RadioAstron VLBI datastream on the ground
station(s), and because of inherent limitations in the data timestamp
information provided by RadioAstron, the timestamp information for the
RadioAstron data are actually set by the tracking station clock at the
moment it begins recording individual VLBI scans.  Modifications to
DiFX therefore also involved changes to the metadata handler in DiFX
and to the delay model server to incorporate not only the relative
time of arrival of the astronomical signal at the spacecraft, but also
to calculate the delay for the transmission of the signal from the
spacecraft to the tracking station.

First fringes obtained with the developed DiFX version, named {\it{dra}}, were obtained on 2012 June
14 using observations of BL~Lac on the RadioAstron--Effelsberg
baseline, and resulting in a clear peak at one location.  %Figure~\ref{Fig:first_fringe} shows a plot of the fringe
%amplitude in delay--rate space, showing a clear peak at one location,
%which matched approximately the solution found by the ASC correlator
%for the same dataset.
Approximately same solution in the delay--rate space was found by the 
ASC correlator for the same dataset.
%
%\begin{figure}
%  \centering
%  \includegraphics[width=0.5\textwidth]{zoom.jpg}
%  \caption{First fringe detection using the space-VLBI enhanced DiFX
%    correlator to process RadioAstron--Effelsberg data of the active
%    galaxy BL~Lac.  The plot shows the fringe intensity (color) as a
%    function of possible delay (horizontal axis) and rate (vertical
%    axis) values in the fringe search window.\label{Fig:first_fringe}}
%\end{figure}
%
\begin{figure}
  \centering
  \includegraphics[width=0.4\textwidth]{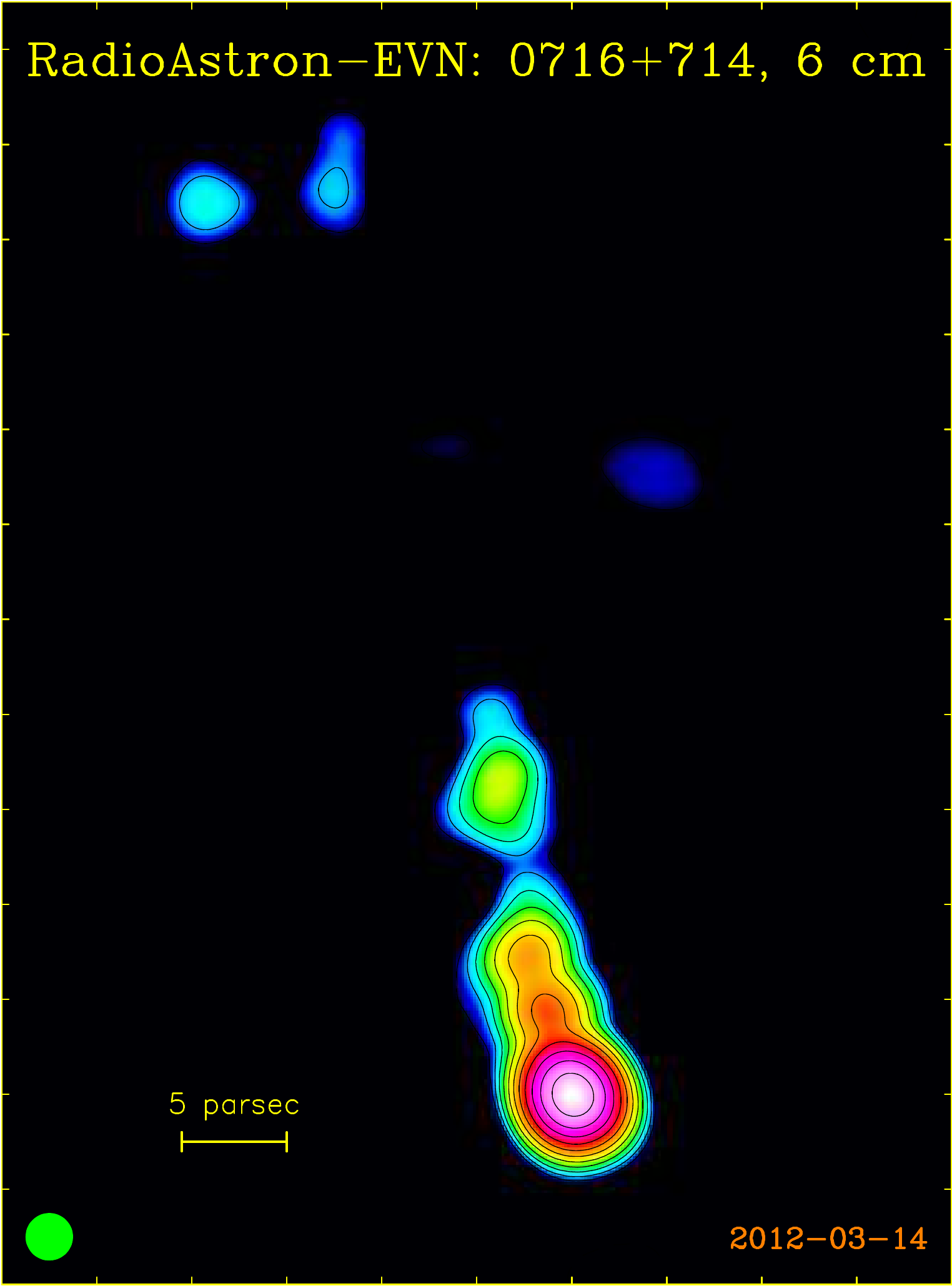}\hfill
  \caption{Image of the active galaxy $0716+714$ at 4.9~GHz made from
    data produced by the {\it{dra}}-DiFX correlation of EVN project EK032C
    observed on 2012 March 14th, including baselines to RadioAstron [2].
    %(Kovalev (on behalf of the RadioAstron mission) 2012, 11th
    %European VLBI Network Symposium \& Users Meeting,
    %submitted). % Right: Image of B$0716+714$ made at 15.4~GHz with
    %approximately the same spatial resolution produced by the
    %MOJAVE Program from an observation on 2012 February 6th
    %(http://www.physics.purdue.edu/MOJAVE/sourcepages/0716+714.shtml).
    $0716+714$ is a BL~Lac-type object that was selected for the
    initial RadioAstron imaging commissioning observations because of
    its compact core, high declination allowing continuous ground
    observation, and sky location matching RadioAstron's observing
    constraints for the time of year.\label{Fig:0716_image}}
\end{figure}
Following the initial fringe result and continued development and
commissioning of the {\it{dra}}-DiFX correlator, it was used to
correlate the first imaging experiment involving RadioAstron. Figure~\ref{Fig:0716_image} shows the initial
imaging result obtained by processing the EK032C experiment [2], using AIPS and DIFMAP for calibration and imaging, respectively.  
%At the time that this dataset was correlated in Bonn using DiFX, the ASC
%correlator output was not able to produce FITS-IDI files that could be
%properly read into AIPS for processing.  (The ASC correlator has
%subsequently been fixed in this regard.)
%
The output datasets from the DiFX software have also been used to make
direct comparisons with the ASC correlator results.  Such comparisons
identified a problem in the ASC $(u,v,w)$ calculations, which has
subsequently been fixed.

\emph{dra}-DiFX is presently used only at the MPIfR, for correlation of RadioAstron data. The merging with the publicly available \emph{dtrunk} version of the DiFX correlator is ongoing. In the following, a summary of the additional capabilities and modifications:

\begin{itemize}

\item RDF-Mark5B conversion routine, to read in data from RadioAstron spaceborne antenna
\item Enabling delay model server Calc (Calc/Solve Package) to calculate delay information for a spaceborne antenna 
\item Introducing general relativistic corrections in the delay model
\item Changing DiFX metadata system to deal with variable position/velocity of the spaceborne antenna
\item Calculating the delay for the transmission of the signal from the spacecraft to the tracking station
\item Calculating the equivalent of parallactic angle correction for the spaceborne antenna from the antenna orientation obtained from the telemetry information.
\item Wide fringe-search windows can be used, thanks to the customized fringe-fitting software.
\end{itemize}
Future improvements will include:
\begin{itemize}
\item Spacecraft acceleration terms correction
\item Inclusion of the on-board maser information from telemetry data
\item Automatic conversion of the native RDF data format to Mark5B
\end{itemize}

%%%%%%%%%%%%%%%%%%%%%%%%%%%%%%%

%This research has made use of data from the MOJAVE database that is maintained by the MOJAVE team (Lister et al., 2009, AJ, 137, 3718).
%%%%%%%%%%%%%%%%%%%%%%%%%%%%%%%
\end{document}